\documentclass[bibyear]{aa}  

\usepackage{graphicx}
\usepackage{txfonts}

\usepackage{url}
\usepackage[pdfborder={0 0 0 },urlcolor=blue,breaklinks]{hyperref}
\hypersetup{colorlinks, citecolor=blue, filecolor=black, linkcolor=black, urlcolor=blue}
\usepackage[normalem]{ulem}	
\usepackage{natbib}
\usepackage{color}

\bibpunct{(}{)}{;}{a}{}{,} 

\newcommand{\BE}{\begin{equation}}
\newcommand{\EE}{\end{equation}}
\newcommand{\BA}{\begin{eqnarray}}
\newcommand{\EA}{\end{eqnarray}}
 \newcommand{\Fig}[1]{Figure~\ref{fig_#1}}
 \newcommand{\Figs}[2]{Figures~\ref{fig_#1} and \ref{fig_#2}}
 
 \newcommand{\Sect}[1]{Section~\ref{sect_#1}}
 \newcommand{\sect}[1]{Section~\ref{sect_#1}}
 \newcommand{\sects}[2]{Sections~\ref{sect_#1} and~\ref{sect_#2}}
 
 \newcommand{\eq}[1]{Eq.~(\ref{eq_#1})}

 \newcommand{\tab}[1]{Table~\ref{tab_#1}}

\newcommand{\rmd}{ {\mathrm d} }

\newcommand{\cf}{\textit{cf.}}
\newcommand{\eg}{\textit{e.g.},}

\newcommand{\ie}{\textit{i.e.}}

\newcommand{\idea}[1]{}
\newcommand{\Bz}{B_{\rm z}}
\newcommand{\Bmax}{B_{\rm max}}
\newcommand{\Bzmin}{B_{\rm z, min}}
\newcommand{\Bzmax}{B_{\rm z, max}}
\newcommand{\PBz}{\mathcal{P}(B_{\rm z})}
\newcommand{\rmax}{r_{max}}

\begin{document} 
		
\title{Evolution of the Magnetic Field Distribution of Active Regions}
	
	\author{S. Dacie
		\inst{1}
		\and
		P. D\'emoulin\inst{2}
		\and
		L. van Driel-Gesztelyi\inst{1, 2, 3}
		\and 
		D. M. Long\inst{1}
		\and
		D. Baker\inst{1}
		\and
		M. Janvier\inst{4}
		\and	
		S. L. Yardley\inst{1}
		\and
		D. P\'erez-Su\'arez\inst{1}			
	}
	
	\institute{Mullard Space Science Laboratory, University College London, Holmbury St. Mary, Surrey, RH5 6NT, UK\\
		\email{sally.dacie.14@ucl.ac.uk}
		\and
		Observatoire de Paris, LESIA, UMR 8109 (CNRS), F-92195 Meudon-Principal Cedex, France
		\and
		Konkoly Observatory of the Hungarian Academy of Sciences, Budapest, Hungary
		\and 
		Institut d'Astrophysique Spatiale, UMR8617, Univ. Paris-Sud-CNRS, Universit\'e Paris-Saclay, B\^atiment 121, 91405 Orsay Cedex, France
	}
	
	\date{Received September 15, 1996; accepted March 16, 1997}
	
\abstract
{}
{Although the temporal evolution of active regions (ARs) is relatively well understood, the processes involved continue to be the subject of investigation. We study how the magnetic field of a series of ARs evolves with time to better characterise how ARs emerge and disperse.}
{We examine the temporal variation in the magnetic field distribution of 37 emerging ARs. A kernel density estimation plot of the field distribution was created on a log-log scale for each AR at each time step.  We found that the central portion of the distribution is typically linear and its slope was used to characterise the evolution of the magnetic field. }
{The slopes were seen to evolve with time, becoming less steep as the fragmented emerging flux coalesces. The slopes reached a maximum value of $\sim -1.5$ just before the time of maximum flux before becoming steeper during the decay phase towards the quiet Sun value of $\sim -3$. This behaviour differs significantly from a classical diffusion model, which produces a slope of $-1$. These results suggest that simple classical diffusion is not responsible for the observed changes in field distribution, but that other processes play a significant role in flux dispersion.}
{We propose that the steep negative slope seen during the late decay phase is due to magnetic flux reprocessing by (super)granular convective cells. }


\maketitle

	
\section{Introduction}
\label{sect_Introduction}
  
  \idea{Evolution in the convective zone}
  The evolution of active regions (ARs) in time is a well studied topic in solar physics \citep[see][]{vanDriel15}, yet the processes are still not fully understood.
ARs result from the emergence of buoyant magnetic flux through the photosphere.
This magnetic flux originates at the tachocline and rises through the convection zone in the form of a magnetic flux tube.
During the rise, plasma drains into the flux tube legs, with conservation of angular momentum causing more plasma to drain into the following leg.
This distorts the shape of the flux tube and causes an increase (decrease) in magnetic pressure of the leading (following) leg, as a result of the total pressure balance. 

  \idea{Crossing the photosphere}
When the flux tube reaches the base of the photosphere, its environment changes dramatically; 
in particular it is no longer buoyant.
Magnetic field accumulates here until the undulatory instability or convective upward motions 
allow fragments of the field to rise and break through the photosphere in a series of small magnetic loops \citep[\eg][ and references therein]{Pariat04}.
The opposite polarities of these loops diverge and the like polarities of many of these small loops coalesce to form strong concentrated spots.
The higher magnetic pressure of the leading flux tube leg means that the leading polarity forms a stronger, more compact spot(s) than the following polarity.
This process of fragmented emergence followed by coalescence has been well observed \citep[\eg][]{Zwaan78, Strous96} and also modelled \citep[\eg][]{Cheung10}.

  \idea{Longterm evolution}
During the emergence phase, the two polarity centres diverge, with their separation reaching a plateau around the time the AR achieves its peak flux. 
This indicates that the flux tube is no longer emerging. 
In addition, convective motions of supergranular cells advect the region's field, breaking it apart.
Other processes may also play a role in the dispersion of the AR.
Moving magnetic features are magnetic flux fragments that are observed to move radially outward from sunspots advected by the moat flow \citep{Harvey73}. They may also contribute to the removal of flux from the spot, although there is not yet conclusive evidence for this \citep{vanDriel15}.
Cancellation of AR flux with the background field also contributes to the removal of magnetic flux from the photosphere, as well as cancellation between the two opposite polarities along the internal polarity inversion line of the AR.
The decay phase of ARs is much longer than the emergence phase and can last for several weeks \citep[\eg][]{Hathaway08} or even months \citep[\eg][]{vanDriel99}, with the weaker following spot decaying much faster than the leading spot.

  \idea{Roadmap}
In this study, we analyse the distribution (probability density) of the vertical component of the photospheric magnetic field (or flux density), and below we refer to it simply as the field distribution.  
This is a new method of characterising the evolution of ARs, 
by looking at changes in the field distribution as the regions evolve.
Our analysis is different from the previous studies which analyse the magnetic flux distributions of photospheric magnetic clusters \citep[\eg][]{Parnell2009, Gosic2014}, 
as we do not cluster the photospheric magnetic field in magnetic entities.
\Sect{theory} describes the theoretical background of emergence, clustering (merging) and diffusion with a focus on the magnetic field distribution expected with these physical processes. 
The data used for the observational study are described in \sect{data_methods}, along with the AR area selection code, which defines the pixels used to calculate the field distribution.
The field distribution plots and their characterisation are explained in \sect{KDE_plots}. Sections \ref{sect_Obs_Temporal} and \ref{sect_Obs_Flux} show the statistical results of the characterisation of 37 ARs.
The characterisation reflects the different evolutionary stages; fragmented emergence, coalescence to form strong sunspots and gradual dispersion of the AR.
Then, in Sections \ref{sect_Obs_Decayed} and \ref{sect_Obs_QS} we explore the dispersing phase of ARs as well as the quiet Sun. We next investigate possible issues present for the derived field distributions in \sect{Obs_Issues}. Finally,
the observational and theoretical results are discussed and compared, allowing conclusions to be drawn in \sect{Conclusions}.

\section{Theory}
\label{sect_theory}

\subsection{Emergence and Clustering} 
\label{sect_th_emergence}

   \idea{Emergence}
Numerical simulations of flux emergence are typically set up with an axial field distributed with a Gaussian profile  \citep{Fan01,Toriumi11}.  
A global Gaussian profile of the two polarities would be expected at the photosphere if the magnetic structure did not change significantly upon emergence. However, as the flux tube reaches the photospheric region it splits into tiny flux tubes, which emerge individually and reconnect with one another \citep{Strous96,Pariat04,Cheung14}. 
Then, even if the axial field did have a Gaussian-like distribution in the convective zone, it is not known how this distribution would be transformed at the photospheric level. The physical process of the emergence phase is too complex to derive an expected magnetic field distribution $\PBz$ where $\Bz$ is the vertical magnetic field component.

   \idea{B distribution}
During the later stages of emergence, small magnetic concentrations of like sign merge to form strong polarities (sunspots).  A simple description of such a flux concentration with axial symmetry is the magnetic field profile: 
    \BE  \label{eq_Bz_multipole}
    \Bz (x,y) = \Bmax ((r/a)^2+1)^{-n/2} \,,
    \EE
with $r^2=x^2+y^2$, where $x,y$ are the two orthogonal horizontal spatial coordinates and $a$ defines a characteristic radius of the polarity.  $\Bz$ is maximum at $r=0$ and decreases as $r^{-n}$ for $r \gg a$.

   \idea{Magnetic source case}
The case $n=3$ corresponds to a magnetic source located at $z=-a$ below the photosphere; it models the linear spreading with height, above $z=-a$, of a thin flux tube located below $z=-a$ \citep[see appendix of][]{Demoulin94}.   Such field distributions were used to model the magnetic field of ARs by using a series of flux concentrations with the parameters (position and intensity) fitted to the observed magnetograms.  This allowed the computation of their coronal magnetic topology.  In particular, the derived intersections with the chromosphere of the computed separatrices were found close to the observed flare ribbon locations showing that magnetic reconnection is responsible for the energy release \citep[\eg][]{Gorbachev88,Mandrini95,Longcope05}.
 
   \idea{Probability of Bz}
Next, we compute the probability $\PBz ~\rmd \Bz$ of having a magnetic field value of $\Bz \pm \rmd \Bz/2$ within the magnetic polarity.    This probability corresponds to the area $2 \pi ~r ~\rmd r$
with $r$ related to $\Bz$ by \eq{Bz_multipole}. Differentiation of \eq{Bz_multipole} provides,
    \BA  
    \rmd \Bz &=& - \Bmax \frac{n}{a^2} ~\frac{1}{((r/a)^2+1)^{1+n/2}}  ~r~\rmd r \nonumber \\   
     &=& - \Bmax \frac{n}{a^2} ~\left( \frac{\Bz}{\Bmax}\right)^{1+2/n} ~r~\rmd r 
            \,. \label{eq_dr_dBz_multipole}
    \EA
By substituting \eq{dr_dBz_multipole} into $\PBz ~|\rmd \Bz| \propto 2 \pi ~r ~|\rmd r|$ one gets 
    \BE  \label{eq_PBz_multipole_step}
    \PBz \propto |\Bz | ^{-1-2/n} \,.
    \EE
The proportionality constant is found by setting the total probability to unity. This integral is calculated in the $r$-range $[0,\rmax ]$ corresponding to the $\Bz$-range $[\Bzmax,\Bzmin ]$. This gives, 
    \BE  \label{eq_PBz_multipole}
    \PBz = \frac{2}{n} \frac{|\Bz | ^{-1-2/n}}{(|\Bzmin |^{-2/n}-|\Bzmax |^{-2/n})} \,.
    \EE
This result extends approximately to a series of flux concentrations modelling an AR as long as the source fields do not overlap significantly. 
For $n=3$, representing a magnetic source, $\PBz \propto |\Bz | ^{-1.67}$, producing a slope of $-1.67$ when plotting $\PBz$ against $\Bz$ on a log-log plot.
At the limit of very large $n$ values $\PBz \propto |\Bz | ^{-1}$, as for the diffusion case of one polarity (\cf~ \sect{th_diffusion_1} and \eq{PBz_one}). Although there is no clear physical interpretation of this, it is unsurprising that differing $\Bz$ distributions, which depend on two spatial dimensions $x$ and $y$, can produce the same $\PBz$, which varies as a function of only one variable, namely $Bz$.

\subsection{Diffusion of a magnetic polarity} 
\label{sect_th_diffusion_1}

   \idea{Classical diffusion}
Already during the emergence, and even more so during the decay phase, the AR magnetic field is affected by convective cells at various spatial scales (\ie\ by granules and supergranules).
The AR magnetic field is progressively dispersed in an ever increasing area nearly proportional to the time duration since the emergence started \citep[\eg][]{vanDriel-Gesztelyi03}.  This is the behaviour expected with classical linear diffusion.  However, the dispersion of an AR also involves local mechanisms not included in classical diffusion, such as clustering of the field at the border of convective cells, cancellation of opposite-sign polarities and submergence of small scale loops.  
Classical diffusion can provide an explanation for the area evolution, but can it also explain the observed magnetic field distribution?

   \idea{Diffusion equation}
The classical diffusion of the vertical field component $\Bz (x,y,t)$ is governed by
    \BE  \label{eq_diffusion}
    \frac{\partial \Bz }{\partial t} = k ~ \left( \frac{\partial^2 \Bz }{\partial^2 x}  
    + \frac{\partial^2 \Bz}{\partial^2 y} \right) \,,
    \EE
where $k$ is a constant coefficient and $t$ is time.

   \idea{One polarity}
For a concentrated initial field with one polarity of magnetic flux $F$, the solution of \eq{diffusion} is:
    \BE  \label{eq_Bz_one}
    \Bz (x,y,t) = \frac{F}{4\pi k~t} \exp \left( -\frac{r^2}{4~k~t} \right) \,,
    \EE
with $r^2=x^2+y^2$.
      
   \idea{Probability of Bz}
As in \sect{th_emergence}, we compute the probability $\PBz ~\rmd \Bz$ of having a magnetic field value of $\Bz \pm \rmd \Bz/2$ within the magnetic polarity, which again corresponds to an area $2 \pi ~r ~\rmd r$.
In this case, $r$ is related to $\Bz$ with \eq{Bz_one}. Differentiation provides,
    \BE  \label{eq_dr_dBz_one}
    2 \pi ~r ~\rmd r = - \frac{4\pi k~t}{\Bz} ~\rmd \Bz \,.
    \EE
Normalisation by the total probability in the $\Bz$-range $[\Bzmax,\Bzmin ]$ gives, 
    \BE  \label{eq_PBz_one}
    \PBz = \frac{1}{|\Bz | ~\ln (\Bzmax/\Bzmin)} \,.
    \EE
Plotting $\PBz$ against $\Bz$ in a log-log plot, produces a slope of $-1$.


\begin{figure}
	\centering
	\includegraphics[scale=0.3]{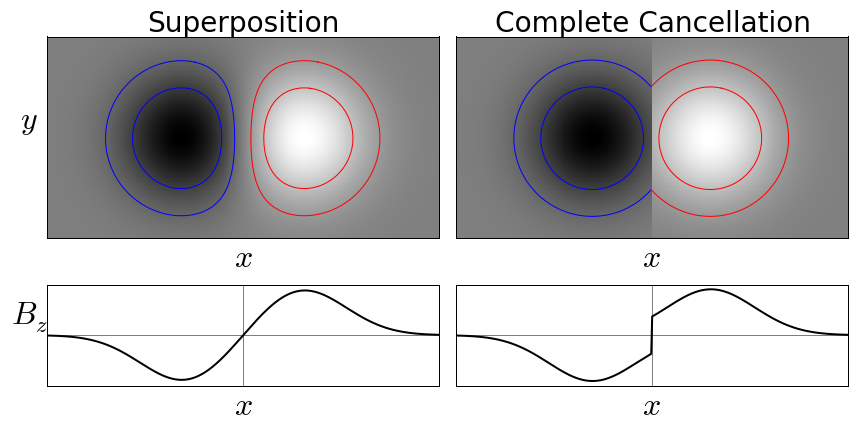}
	\caption{The two different scenarios for flux dispersion of a bipole as described in \sect{th_diffusion_2} are shown here. The left hand side of the plot shows superposition with no cancellation and the right hand side shows the case where complete cancellation occurs at $x=0$. Two isocontours are also shown for both the negative and positive spots. The two plots in the bottom row show a cross section of the magnetic field values taken at $y=0$.} 
	\label{fig_sup_cancellation}
\end{figure}

\begin{figure}
	\centering
	\includegraphics[width=0.48\textwidth]{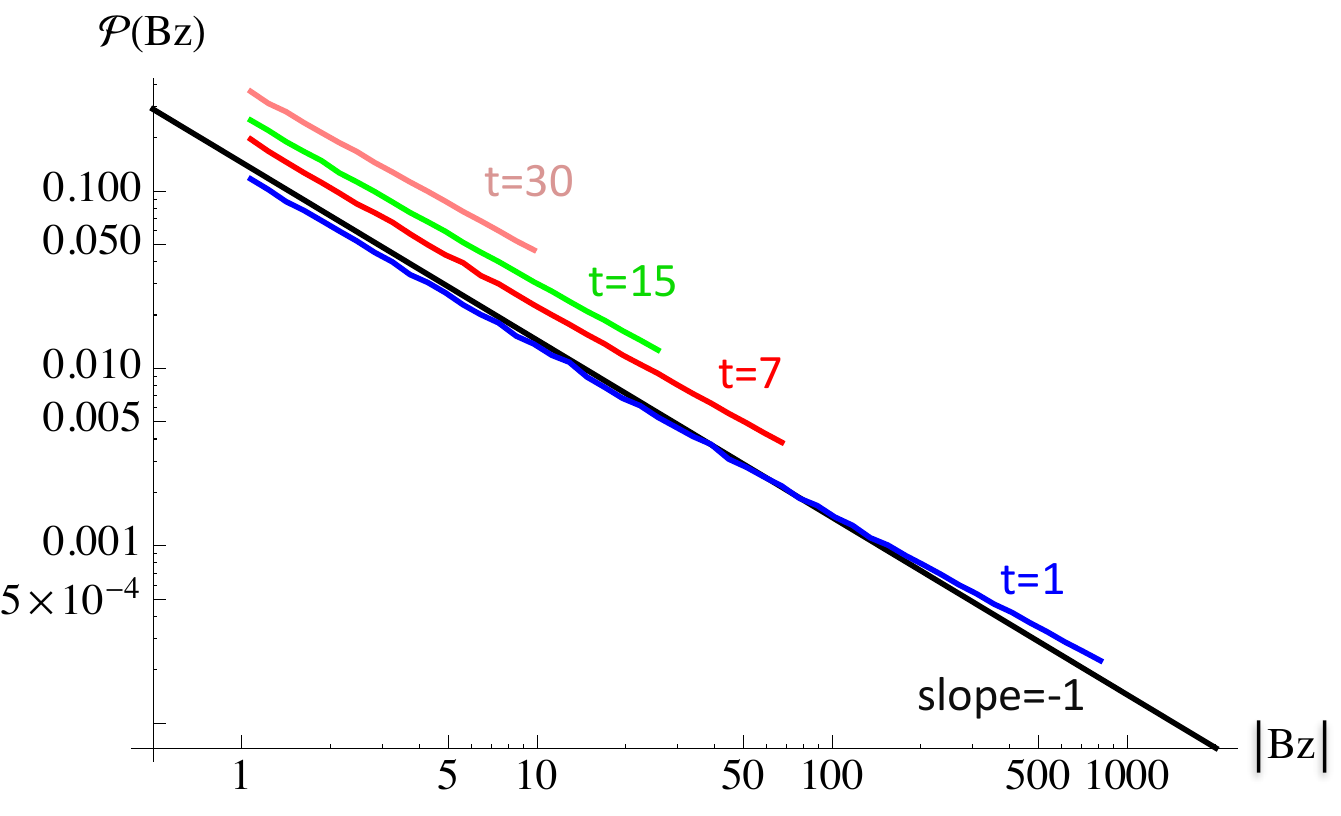}
	\caption{Log-log plot of the probability distribution of $\Bz$ versus $\Bz$ for a bipolar field diffusing with time. The $\Bz$ spatial distribution is described by \eq{Bz_two}. The time $t$ is normalised by the initial time when $R/a=1$ is selected (blue line).  Only $\Bz>1$ is supposed to be detectable.  The black straight line has a slope $-1$ for reference.  }
	\label{fig_dist_diffusion}
\end{figure}

\subsection{Diffusion of a magnetic bipole} 
\label{sect_th_diffusion_2}

	\idea{Bipole}
In this subsection we consider the evolution of a simple bipole to model the global evolution of an AR. We suppose that the centre of the two polarities are located at the fixed positions $x=\pm R, ~y=0 $.
Since \eq{diffusion} is linear in $\Bz $, the evolution of the bipole is the superposition of two solutions of \eq{Bz_one} when shifted spatially by $\pm R$:
\BE  \label{eq_Bz_two}
\Bz (x,y,t) = \frac{F}{4\pi k~t} 
\left( e^{-\frac{(x-R)^2}{4~k~t}}-e^{-\frac{(x+R)^2}{4~k~t}} \right)
e^{-\frac{y^2}{4~k~t}} \,,
\EE
where 
$F$ is the total flux of the isolated positive polarity.
This is illustrated in the top left panel of \Fig{sup_cancellation}, with the bottom left panel showing a cross section of the $\Bz $ values taken at $y=0$.
   
	\idea{Describe the magnetogram evolution}
The spatial distribution of $\Bz$ in \eq{Bz_two} depends on the parameters $F$, $a=\sqrt{4~k~t}$ and $R$.  $F$ defines the field strength while $a$ defines the polarity size and both are scaling factors.  The main parameter of \eq{Bz_two} is $R/a$. For $R/a \gg 1$ the polarities are well separated and do not overlap, while for $R/a \ll 1$ the opposite polarities mostly cancel each other leaving a dipolar magnetic field.
   
	\idea{Probability of Bz}
The probability distribution of $\Bz$ is computed numerically since the $\Bz$ isocontours are not simply circles ($\PBz$ is computed by summation along these isocontours). In a log-log plot, the slope is in the range $[-1,-0.94]$ at all times during the diffusive evolution (\Fig{dist_diffusion}). More precisely, the slope $\approx -1$ for $R/a>3$ and converges rapidly to $-1$ when the separation between the polarities is larger, as expected from \eq{PBz_one}. The slope is maximum, $\approx -0.94$, for $1 \leq R/a \leq 1.7$, while it shows a slight decrease to $\approx -0.96$ for lower $R/a$ values. When the polarities are interacting
the slope is slightly less steep than with one polarity (slope $=-1$) because the other source acts to decrease $\Bz$ and even removes weak values surrounding the inversion line.   This implies fewer cases with a given $\Bz$ value, particularly for low $\Bz$, making the global slope of the distribution slightly less steep.  Finally, the increase of probability values with time, as seen in \Fig{dist_diffusion}, is simply due to the normalisation of the total probability to 1 while the maximum of $\Bz$ is decreasing with time and the minimum of $\Bz$ is selected to be fixed (here to $\Bz =1$), modelling an instrumental detectability threshold.  

   \idea{Flux evolution}
The net flux evolution is computed by performing the integration on the positive polarity, where $x>0$, which gives, 
    \BE  \label{eq_F_two}
    F(t) = F(0) ~{\rm erf} (R/\sqrt{4~k~t}) \,,
    \EE
where ${\rm erf}$ is the error function ${\rm erf}(x)= 2/\sqrt{\pi} \int_{0}^{x} e^{-u^2} \rmd u$. For small diffusion times, $\sqrt{4~k~t} \ll R$, $F(t)$ is nearly constant, while  for large diffusion times \eq{F_two} can be approximated as  $F(t) \approx F(0) ~R/\sqrt{{\pi}k~t}$ so that the flux of the polarities decreases as $t^{-1/2}$.  The cancellation of flux is due to the superposition of the opposite polarities in the $x$ direction.  This is a 1D process, so it scales as the ratio of the sizes in the $x$ direction: $R/\sqrt{4~k~t}$.

   \idea{Immediate cancellation}
Equation (\ref{eq_Bz_two}) describes the diffusion of each polarity isotropically, \ie, they do not cancel but just superpose spatially.  This implies an apparent loss of magnetic flux by superposition of opposite $\Bz$ values when computing the total flux, however there is not a physical cancellation since the full amount of both polarities continue to diffuse with time.  The other extreme behaviour, a complete cancellation at the inversion line, can be described by the field:
    \BE  \label{eq_Bz_cancellation}
    \Bz (x,y,t) = \frac{F}{4\pi k~t} 
          ~{\rm ext} \left( e^{-\frac{(x-R)^2}{4~k~t}},-e^{-\frac{(x+R)^2}{4~k~t}} \right)
                 ~e^{-\frac{y^2}{4~k~t}} \,,
    \EE
where the function
ext$(a,b)$ is defined as $a$ if $|a|>|b|$, $b$ if $|a|<|b|$, and $0$ otherwise (then it is a signed extremum function except at $|a|=|b|$ where it vanishes to model immediate cancellation).
This scenario is shown in the top right panel of \Fig{sup_cancellation}, with the bottom right panel showing the $\Bz$ values at $y=0$.

   \idea{Physics of diffusion/cancellation in the photosphere}
Equation (\ref{eq_Bz_cancellation}) describes two magnetic polarities which are diffusing independently of each other (no overlap of $\Bz$ as in \eq{Bz_two}) and they exactly cancel at $x=0$ (no diffusion in the region of the other polarity).  In the photosphere the opposite polarities diffuse into each other as a result of advection by convective cells of small flux tubes, which at some point meet and cancel.  This process is mostly driven by the supergranular flows and the typical distance of travel across the inversion line in the other polarity  (the mean free path) is expected to be of the order of a supergranular cell (about 30 Mm).  Equations~(\ref{eq_Bz_two}) and (\ref{eq_Bz_cancellation}) are the limits when the mean free path is infinite and zero, respectively.

   \idea{Probability of Bz}
The probability of $\Bz$, for values where the corresponding isocontour is not touching the inversion line ($x=0$), given by \eq{Bz_cancellation}, has the same $\Bz$ dependence as \eq{PBz_one}. 
For the lower $\Bz$ values, some of the pixels are not present (those which would be located on the other side of the y axis if they had not undergone cancellation) 
so the probability is slightly lower.
In summary, the distribution $\PBz$ shown in a log-log plot (as in \Fig{dist_diffusion}) has a slope of exactly $-1$ for large $|\Bz|$ values and slightly above $-1$ (less steep) for low $|\Bz|$ values.

   \idea{Summary}
All in all, this implies that $\PBz$ is similar for an inefficient cancellation rate (\eq{Bz_two}) and for a very efficient one (\eq{Bz_cancellation}).  Thus, any cancellation rate, with a classical diffusion of the polarities is expected to imply a slope $\gtrsim -1$ for $\PBz$ drawn in a log-log plot.

\begin{figure}
	\centering
	\includegraphics[scale=0.4]{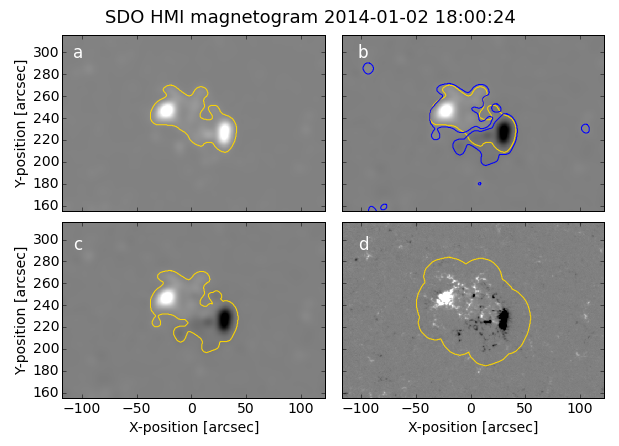}
	\caption{Example of the area selection and dilation procedure (\sect{data_AR_extension}) for NOAA AR 11945.
		Panel a shows the selection $>$ 40 G in the blurred map of absolute field values.
		Panel b shows the selection	$>$ 20 G and $<$ -20 G in the smoothed signed map (blue), along with the selection from panel a (yellow).
		Panel c shows the selection	after the field dependent dilation and panel d the final region selection after the field independent dilation.}
	\label{fig_area_selection}
\end{figure}

\section{Data and AR Selection}
\label{sect_data_methods}

\subsection{Data and data treatment} 
\label{sect_data_AR_Magnetograms}

   \idea{Data set}
We studied the magnetic field evolution of 37 ARs that emerged on the solar disk between June 2010 and the end of 2014.
To ensure consistency and that the emergence time could be correctly identified, only those ARs that emerged in relatively magnetic-field free regions (without detectable remnant of other ARs) were chosen. In addition, only ARs emerging prior to central meridian passage (CMP) were selected to analyse all, or at least most, of the emerging phase.
The resulting ARs are given in \tab{Max}.

   \idea{Selection of ARs}
The ARs were studied using photospheric line-of-sight magnetograms from the Helioseismic Magnetic Imager \citep[HMI;][]{Schou12} on board the \textit{Solar Dynamics Observatory} \citep[SDO;][]{Pesnell12}.
The time at which each AR started to emerge was found by visual inspection of the line-of-sight magnetograms.  Although this method is subjective and is influenced by the position of the AR on the solar disk \citep[issues which are discussed in][]{Fu16}, these effects were negligible as a result of the relatively low time cadence data (6 hours) used here to study the ARs.

For the decaying ARs discussed in \sect{Obs_Decayed}, the ARs began to emerge on the far side of the Sun. In order to estimate the emergence start times in these cases, we used data from the Extreme UltraViolet Imager (EUVI) telescope onboard the \textit{Solar Terrestrial Relations Observatory Behind} \citep[STEREO B;][]{Kaiser08} spacecraft.

   \idea{B corrections}
The magnetic field data were first treated using a cosine correction assuming that the magnetic field is locally vertical.  
This was followed by a derotation to central meridian to correct for area foreshortening.

\subsection{Definition of AR area} 
\label{sect_data_AR_extension}

   \idea{Intro: define AR extent}
A rectangular submap surrounding the maximum extent of the AR was manually defined, with the area of the AR itself defined using a semi-automated technique \citep[adapted from that outlined in][]{Yardley16}.
This process was important to ensure an unbiased estimation of the field distribution of the AR, especially as the rectangular submap contains magnetic field from the quiet Sun as well as possibly flux from neighbouring regions which would affect the results.

   \idea{Initial area selection}
To focus on the larger scale features rather than individual flux fragments, the absolute values of the magnetic field within the submap were smoothed using a Gaussian filter with a standard deviation (width) of 7 pixels.
The AR extension is then automatically defined from the areas in which the pixels have smoothed values $>$ 40 G (shown in \Fig{area_selection}a).
For the initial stages of flux emergence, our code searches for a bipolar region, defined as having between a fifth and five times the amount of positive flux as negative flux.
If no bipole is found a search is made for positive and negative bubbles that are close to each other (within a distance of 20 pixels, approximately 10 arcseconds).
To include regions of fragmented flux emergence, more of the original bubbles are accepted if they lie within 20 pixels of this selection or 10 pixels of these additional bubbles. 

   \idea{Later area selection}
As ARs evolve, they change significantly, with the sea serpent structures coalescing into two main polarities which grow to cover a larger area of the solar surface.
	To reflect these changes, the AR area selection criteria change
after the unsigned flux of the near bipole region has reached $8 \times 10\textsuperscript{20}$ Mx, when the AR area was found to dominate the rectangular submap.
The AR area is still chosen from the regions of pixels with smoothed values $>$ 40 G, but the selection criteria are different, with larger regions and regions close to the large regions being chosen.
To capture fragments of flux that break away from the main polarities, this area is then extended to neighbouring regions (within 10 pixels).

   \idea{Checking}
If at any time step the automated selection is not satisfactory, for example if there are inconsistencies between neighbouring time steps (\ie ~flux at the edge of the region is not included in the selection at one time step, but is included in both the preceding and the following time steps), the selection can be made manually from the smoothed submap.

%
%

   \idea{Dilations}
Once the region has been selected, dilations are applied to include the region immediately surrounding the selection, as this was found to have a different magnetic field distribution to that of the quiet Sun (see \sect{Obs_Issues}) and is therefore assumed to be strongly influenced by the emerging flux.
For the first dilation, the signed data were smoothed using a Gaussian filter of width 7 pixels.  The pixels of smoothed values $>$ 20 G or $<$ -20 G are selected (shown in \Fig{area_selection}b with blue contours). 
Each of these regions is compared to the previously selected region (yellow contour), and if there is some overlap, the pixels in that region are unmasked (panel c of \Fig{area_selection}). 
Because the blurring is applied to the signed values, the dispersing parts of the region are likely to be accepted, while any diffuse neighbouring regions of opposite polarity should be avoided.
Finally two field-independent dilations are applied, both using a disk-shaped kernel with a radius of 12 pixels (providing a smoother edge than a single dilation with a larger kernel).
All the pixels in the dilated region (shown \Fig{area_selection}d) are taken for the analysis.

\section{Kernel Density Estimation Analysis}

\label{sect_KDE_plots}

\subsection{Method} 
\label{sect_KDE_Method}

  \idea{Introduce KDE plots / histograms}
Kernel density estimation plots \citep[KDEs;][and references therein]{Wegman72, Silverman86} were created for each time step, showing the distribution of pixels with respect to field value (flux density) on a log-log scale. 
KDEs are very similar to histograms, in that both types of plots show the distribution of data points.
However, histograms rely on discretising the data points, using user-defined bins, which normally leads to a loss of information. Moreover, the arbitrary choice of bin positions can affect the subsequent interpretation.
On the other hand, KDEs assign a kernel of a certain width to each data point, effectively smoothing the finite number of data points.
This KDE technique has been shown to be superior to histograms both theoretically and in applications \citep[\eg][]{deJager86,Vio94}, and KDEs have been used in a broad range of astrophysics domains \citep[\eg][]{SchulzeHartung12, Sui14}.

\begin{figure}
	\centering
	\includegraphics[scale=0.5]{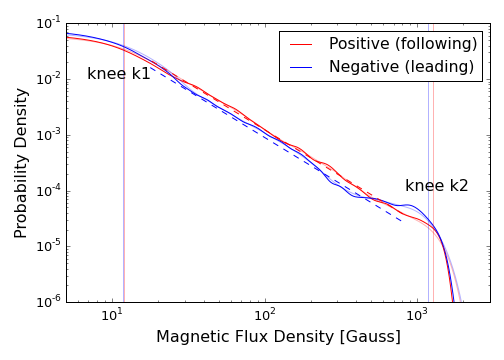}
	\caption{The field distribution of NOAA AR 11945 18 hours after the start of its emergence. The distribution is calculated from the pixels within the contour shown in panel d of \Fig{area_selection}. The positions of the knees, k1 and k2 (vertical lines), and the lines of best fit to the middle section (dashed) are shown for both polarities.}
	\label{fig_example_kde}
\end{figure}

  \idea{Managing KDE plots}
The applied kernels of KDE plots can take on any shape and were chosen to peak at the data point value and to decay away from this.
The kernel widths can also be varied depending on the data point value. 
A larger width can be used where there is less data, enabling the noise in the distribution to be made roughly constant.
While the smoothness of a KDE clearly depends on the kernel widths, similar to the dependence of histograms on their bin widths, unlike histograms there is no dependence on an arbitrary choice of bin position (the kernel is applied and centered on each data point).

  \idea{Application to this study}
The KDEs in this study\footnote{The code used to create and analyse KDEs in this study is available here: https://github.com/SallyDa/Sally\_KDE}
 were created using a Gaussian kernel with standard deviation equal to 2.5 G for field values between -25 and +25 G and equal to $|B_z|/10$ for $|B_z|> 25$~G.
This was chosen to give a plot with the noise level being roughly independent of $B_z$ (as it would appear on a log-log plot). 
From the theoretical considerations of \sect{theory}, a line with a slope of $-1$ would be expected for the KDE when plotted on a log-log scale. 
This means that the pixel frequency would be proportional to $1/|B_z|$. 
The optimal KDE should be built with a variable kernel width such that the kernel width is inversely proportional to the frequency at that point. 
Thus a kernel width proportional to $|B_z|$ was used. 
There are a few reasons why a lower limit on the kernel width was chosen for low $|B_z|$ values. 
One reason is to represent the instrumental errors, which reduce the precision of the measurement. 
Another reason is that the frequency does not increase indefinitely as the value 
$|B_z|$ decreases. 
Instead, it is expected that the distribution and its derivative are continuous across 0 Gauss on a linear $B_z$ scale. 
Indeed the KDE plots do show that
the distributions are flatter for $|B_z|$ values less than $\sim$ 25 G (\eg\ \Fig{example_kde}).

\subsection{Observed Distributions} 
\label{sect_KDE_Distributions}

  \idea{Characteristics of B distributions}
All the KDEs were found to have some common features, namely a flatter section below $\sim$ $10$~G 
followed by a turning point (the first knee, at field value k1, indicated by a vertical line on \Fig{example_kde}), a steeper section up to $\sim$ 1000 G and then another turning point (the second knee, at field value k2, also indicated by a vertical line) after which the frequency drops off rapidly.
Although k2 varies greatly with the age and strength of the AR, it was not used to characterise the distribution as it is highly dependent on only a very small proportion of pixels.

  \idea{Slope}
To capture only the larger scale features, an extra smoothed KDE plot was created using Gaussian kernels of twice the width as before, and the positions of the knees (k1 and k2) were found from this.
k1 was taken as the point of most negative second derivative before the slope goes to values less than $-1$.  
k2 was taken as the first time, moving along the KDE graph from right to left, that the slope becomes more positive (less steep) than $-3$.
A best fit line was fitted to the original (not extra smooth) KDE plot between $1.5 \times$ k1 and $\frac{2}{3} \times$ k2. 
The best fit line was calculated using the method of least squares, taking data points evenly spaced along the log(B) axis. 
Examples are shown (dashed lines) in \Fig{example_kde} for both the positive and negative distributions.

\begin{figure*}
	\centering
	\includegraphics[scale=0.45]{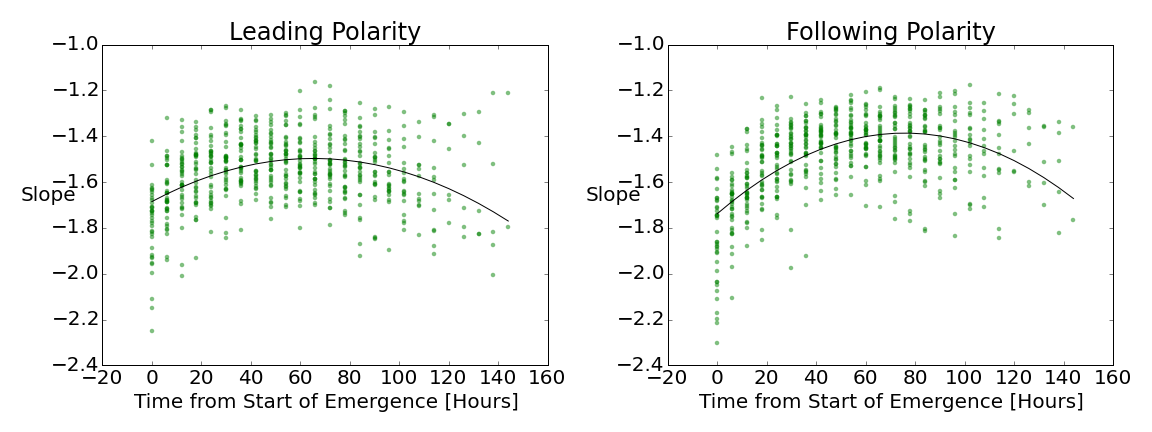}
	\caption{The slope of the best fit straight line to the middle section of the log-log field distribution plot was calculated as illustrated in \Fig{example_kde}. This plot shows each slope value plotted against the age of the AR from the start of its emergence ($t=0$) at each time step and for each AR. Data for leading and following sunspots were separated. Second order polynomials were least-square fitted, showing the general trend.}
	\label{fig_gradvstime}
\end{figure*}

\begin{figure*}
    \sidecaption
	\centering
	\includegraphics[scale=0.42]{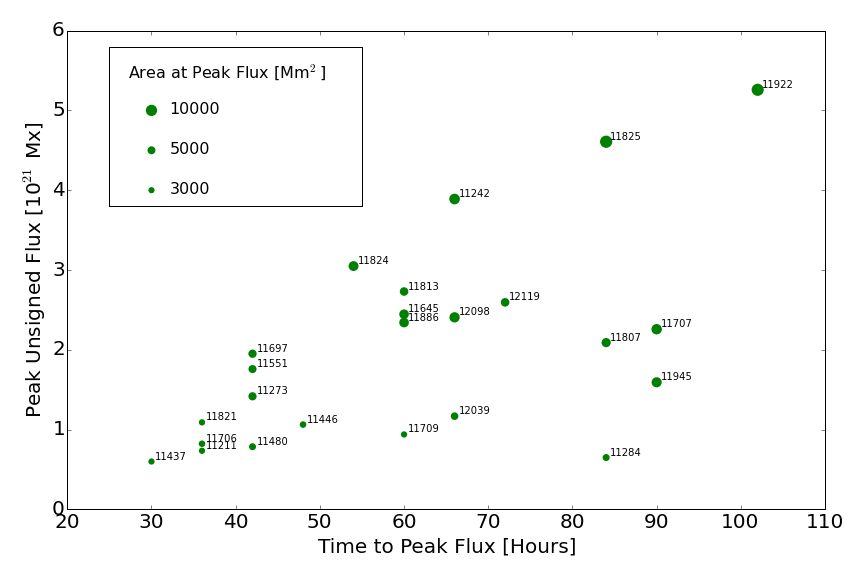} 
	\caption{The peak unsigned flux reached by each AR is plotted against the time it took to reach that value from the start of the region's emergence. The size of the points gives an indication of the area covered by the region at that time. The points are labelled by their NOAA AR number.
 }
	\label{fig_flux_vs_time}
\end{figure*}

\begin{figure*}
	\centering
	\includegraphics[scale=0.45]{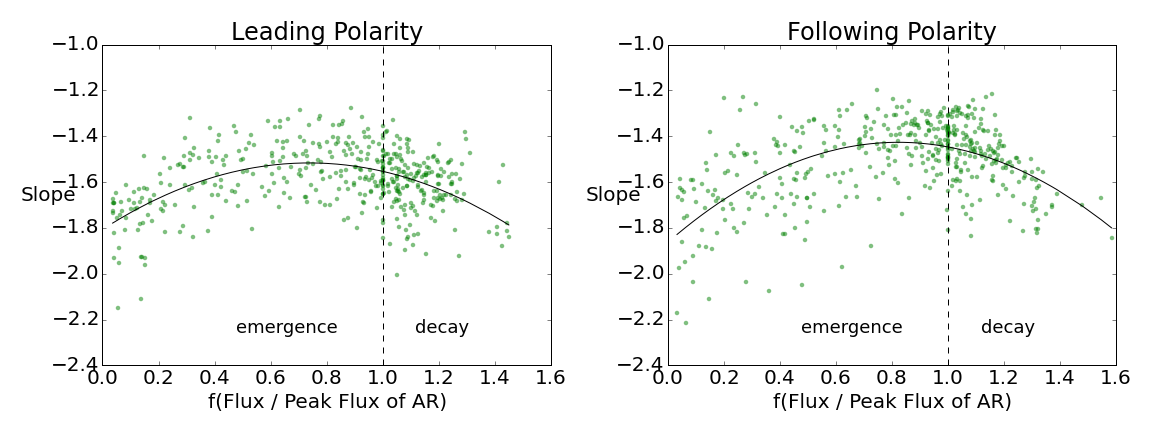}
	\caption{Similar to \Fig{gradvstime}, this plot shows the evolution of the slope. Here, the slope is plotted against a function of the normalised magnetic flux. This normalised flux is defined as the ratio of the AR flux to the peak flux that region achieves. The peak flux and the ratios are found separately for the positive and negative magnetic polarities. To distinguish between emergence and decay phases, the function $f(F)$, defined by \eq{flux}, is used in abscissa. The emergence phase is in the range $[0,1]$ and the decay phase is for abscissa $>1$.
Second order polynomials were least-square fitted, showing the general trend. 
}
	\label{fig_gradvsf(flux)}
\end{figure*}

\section{Observational Results}
\label{sect_Observational}

\subsection{Temporal Evolution} 
\label{sect_Obs_Temporal}

  \idea{Describe slope versus time}
\Fig{gradvstime} shows how the slope of the distributions changes with time from the start of AR emergence for all studied ARs (listed in \tab{Max}).
The slopes start off steep and negative, in about the range $[-2.2,-1.5]$, peak in the range $[-1.7,-1.2]$, and then become steeper (more negative) again as the ARs decay.

  \idea{Interpretation of the results}
The initially steep slope is related to the fragmented emergence.  It is followed by the coalescence of flux elements to form stronger flux concentrations and indeed the slopes around $t=60$ hours are 
comparable to the slopes found for a magnetic source with $n=3$ ($\approx -1.67$) as derived in \sect{th_emergence}, \eq{PBz_multipole}.
The change in slope during the regions' decay is due to dispersion with the number of higher-field pixels decreasing and the proportion of low to middle field pixels increasing.
The latter evolution goes away from the slope given by classical diffusion ($\approx -1$, \sect{th_diffusion_2}).

\begin{figure*}
	\centering
	\includegraphics[scale=0.5]{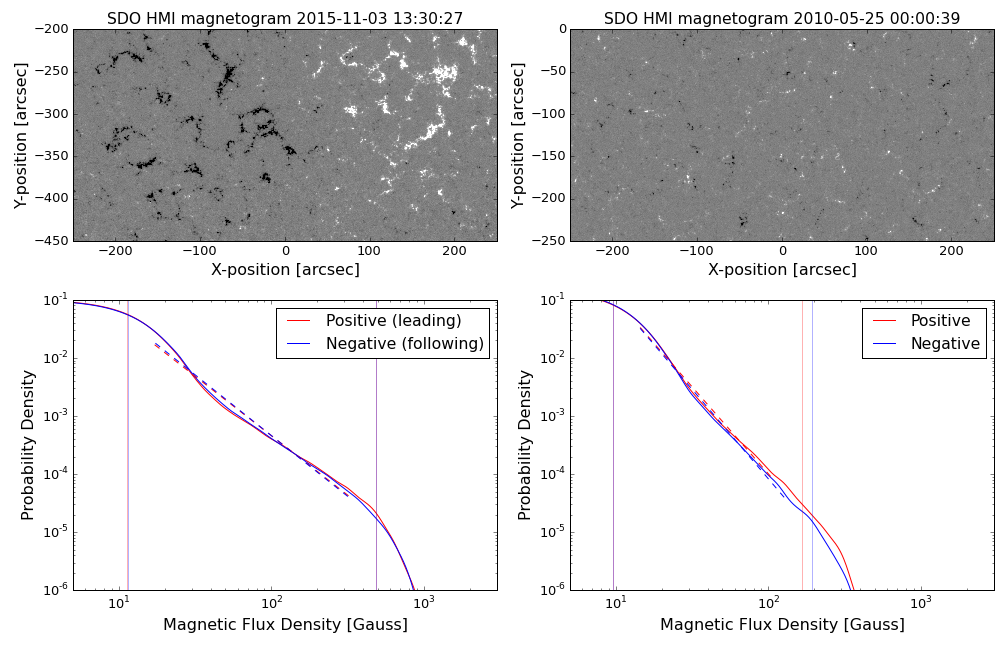}
	\caption{A decayed AR (left column) and a region of quiet Sun (right column) are shown in the top panels, with their magnetic field distributions in the bottom left and right panels respectively. The positions of the knees, k1 and k2 (vertical lines), and the lines of best fit to the middle section (dashed) are shown for both polarities.}
	\label{fig_decaying_region}
\end{figure*}

\subsection{Evolution versus Magnetic Flux} 
\label{sect_Obs_Flux}

  \idea{Why using magnetic flux?}
The timescale of emergence varies broadly from one AR to another \citep[\eg][ and references therein]{vanDriel15}.
As a result, in \Fig{gradvstime} ARs with various flux values and emergence durations are presented together, mixing different phases and field strength distributions of AR evolution (\eg\ $\sim$ 80 hours after the start of flux emergence some ARs are still emerging while smaller ARs are already in their decay phase). 
In this Section, we characterise the emergence stage of an AR by the amount of emerged magnetic flux, normalised to its maximum value.

  \idea{Max flux}
The maximum flux, $F_{\rm max}$, is defined as the maximum unsigned flux achieved during the AR evolution, so long as this value was not reached at the last time step.
24 regions from the original 37 reached peak flux and were included in this part of the study.
The other regions were still in their emergence phase at the end of the observational time period.

    \idea{Max flux versus emergence duration}
\Fig{flux_vs_time} shows that larger ARs generally do take a longer time to emerge, but that the relationship between emergence time and peak flux is not a simple one, with some regions taking a long time to emerge without reaching a particularly high peak flux.
The rates of flux emergence were seen to vary, both during the emergence of single regions and from one AR to another in agreement with the results of \citet{Poisson15,Poisson16} obtained on other sets of ARs.
Some regions were seen to have an initial gradual emergence phase followed by a more rapid phase, while others emerged rapidly at the start and had a decreasing emergence rate later on.

    \idea{Slope versus flux}
The ratio of flux to peak magnetic flux is used in \Fig{gradvsf(flux)} to study the emergence independent of the time it takes. Here the maximum flux is computed separately for the two magnetic polarities. Moreover, in order to separate the emergence from the decay phase (which are in the same range of flux) we use the function $f(F)$ defined as
  \begin{eqnarray}
  f(F/F_{\rm max}) &=& F/F_{\rm max} \qquad \,\,\, \mbox{\rm for } t\leq t_{\rm max}  \label{eq_flux}\\
                   &=& 2-F/F_{\rm max} \quad \mbox{\rm for } t>t_{\rm max}  \,, \nonumber 
  \end{eqnarray}
where $t$ is the time since the start of emergence, $F_{\rm max}$ is the maximum flux and $t_{\rm max}$ the duration of the emergence until $F_{\rm max}$ is reached.
Despite the issues arising due to variable emergence rates, \Fig{gradvsf(flux)} shows an improvement to \Fig{gradvstime} in that the trends, particularly in the case of the leading spot, are more pronounced.

    \idea{Difference between leading and following polarities}
On average, there is an imbalance present in the slopes of the field distribution between the leading and following polarities.
At the beginning of the emergence phase the field distribution slope is slightly more negative for the following than the leading polarity, becoming less negative as the peak value of flux is reached. However, these differences, best seen in the fitted polynomials, are comparable to the standard deviation of the residuals. 
To reduce the statistical noise and investigate whether these slopes are dependent upon parameters other than the normalised flux a larger sample of regions would need to be studied.

    \idea{Summary of the results and interpretation}
In summary, \Fig{gradvsf(flux)} shows one important aspect of the evolution of the magnetic field distribution, namely the transformation of a distribution mainly dominated by the weak fields (steeper slope), at the beginning of emergence, to one with more numerous stronger field pixels, around the maximum flux, and then back to a distribution dominated by weak field pixels.   The typical slopes around the maximum flux correspond to the slope found with the simple model described in \sect{th_emergence}.  In contrast, neither the initial emergence nor the decay phase conforms with the plausible theoretical model we considered.
In particular the distribution during the beginning of the decay shows an evolution opposite to that expected from classical diffusion (with the slope becoming steeper rather than tending towards the expected value $-1$ as found in \sect{th_diffusion_2}).

\begin{figure}
\centering
\includegraphics[scale=0.5]{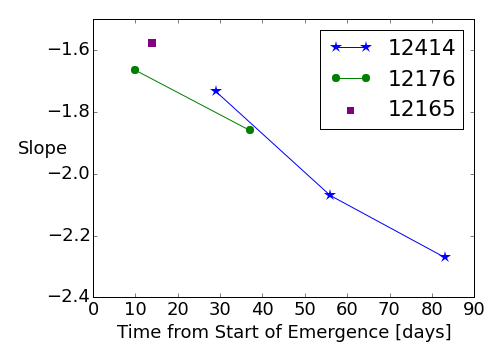}
\caption{The slopes for older decaying ARs are shown against time from the start of the AR's emergence. A downward trend can be seen.
}
\label{fig_slope_vs_days}
\end{figure}

\begin{figure*}
	\centering
	\includegraphics[scale=0.45]{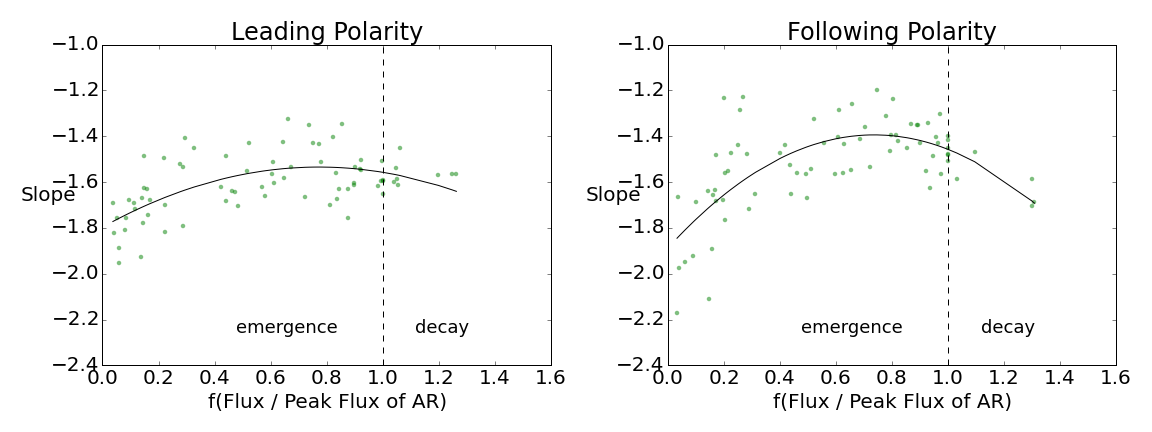}
	\caption{Similar to \Fig{gradvsf(flux)}, but in this case only using data points from ARs within 10 hours, $\sim$ 6 degrees, of their central meridian passage. To distinguish between emergence and decay phases, the function $f(F)$, defined by \eq{flux}, is used in abscissa.
Second order polynomials were least-square fitted, showing the general trend.
	}
	\label{fig_gradnearcm}
\end{figure*}

\subsection{Decayed ARs} 
\label{sect_Obs_Decayed}

The ARs selected for this study were ones that emerged on the solar disk and were tracked until they rotated too far ($\geq 60^{o}$) from the central meridian. As such, all the data came from relatively young ARs, up to just 6 days old.
A preliminary study of decaying ARs was performed to see if the decreasing slope trend continues as ARs continue to age. 
Three relatively isolated decaying ARs (NOAA 12165, 12176 and 12414), in which no new flux emergence was observed, were selected.
The ARs were chosen to be of approximately the same maximum flux ($\sim 5 \times 10\textsuperscript{21}$ Mx), as this affects how quickly the ARs decay and how quickly their distribution changes.
Two of these ARs started emerging on the far side of the Sun, so their emergence start times were identified by eye using 171\,\AA\ data from STEREO B.
Two of the regions were observed for multiple solar rotations, providing data points at later times.

To calculate the field distributions of these regions, all of the pixels within a rectangular subplot centred on the AR were used. This simple procedure was used, as at these times the ARs are quite dispersed and difficult to distinguish from the quiet Sun field. 
An example of one of the decaying ARs and the rectangular subplot that defined the region is shown in the top left panel of \Fig{decaying_region}, with the AR's field distribution shown in the bottom left panel.
The slopes of the three decaying ARs at the different times since they started emerging are shown in \Fig{slope_vs_days}.
Here, an average of the positive and negative slope values was used, as the slopes for the two different polarities were consistently very similar, in each case differing by less than 0.1.
This may not be the case for an AR where the leading polarity retains a coherent spot for longer than the following polarity.

The results shown in \Fig{slope_vs_days} indicate that the downward (steepening) trend of the slope during AR decay continues as the regions continue to age.
A deeper study is needed to confirm this result and to find the dependence on maximum flux.

\subsection{Quiet Sun} 
\label{sect_Obs_QS}

At the end stages of AR decay, the dispersing magnetic field becomes part of the quiet Sun.
We studied four regions to find the slope of the quiet Sun field distribution.
As for the decaying ARs in \Sect{Obs_Decayed}, all the pixels within a selected rectangular area were used to build the field distribution.
One of the four regions and its distribution is shown as an example in the right column of \Fig{decaying_region}.
The regions studied were taken at central meridian on 2010-May-25 (shown in \Fig{decaying_region}), 2015-Nov-25, 2015-Dec-22 and 2016-Jan-18. The latter three regions were centered on -75 arcseconds in Solar Y.

The average of the positive and negative distribution slopes were $-3.0$, $-3.0$, $-2.9$ and $-3.3$, with the difference between the positive and negative being less than $0.3$ for each region. Combining this result with that of \Fig{slope_vs_days}, it implies that the slope of the AR field distribution continues to decrease (steepen) as the AR decays, until reaching the quiet Sun value $\sim -3$. 

A possible origin of this $\sim -3$ slope is the advection of the magnetic field by the diverging flow of supergranules, as follows.  With an axisymmetric magnetic field and purely radial flow at speed $u_r$ away from the cell centre, the ideal induction equation in the stationary regime implies that $B_z(r) ~\propto~ 1/(r~u_r)$ with $r$ being the radial distance away from the cell centre.   
Supposing that $u_r$ is nearly independent of $r$ away from the cell centre and boundary, implies that $B_z \propto 1/r$. Comparing this to \eq{Bz_multipole} provides $n \approx 1$ for $r \gg a$, which produces a slope of $-3$ from \eq{PBz_multipole}.  This rough approach assumes a very simple form for the convective flow and does not include other important physical processes such as the field concentration/cancellation at the cell boundary.  A detailed analysis of observations and/or numerical simulations will be needed to test if the $\sim -3$ slope found in the quiet Sun is mainly due to the advection of the magnetic field by the diverging flow of supergranules.

\subsection{Possible Issues for the Derived Distributions} 
\label{sect_Obs_Issues}

In this section we report some of the tests done to ensure that the derived slopes in emerging ARs are not affected by the inclusion of surrounding quiet Sun or by projection effects.

    \idea{AR borders}
With regard to the area selection process (as described in \sect{data_AR_extension}), we analysed the area bordering the ARs by selecting a ribbon-like region around the ARs and by studying the field distributions for various ribbon thickness. In the bordering region we found a distribution that differed from that of the quiet Sun. 
As such, we decided to include these neighbouring pixels in the selection by applying dilations.
Care was taken to ensure that the dilations were not too large, avoiding the inclusion of decaying field from old ARs nearby, which could be identified by eye.

    \idea{Behavior of the low B: shift k1}
The quiet Sun region was also used to study possible effects arising from the position of the region on the solar disk with respect to distance from the central meridian since we are using the line-of-sight magnetic field data.
Contrary to ARs, the distribution of the quiet Sun is expected not to evolve on a time scale of weeks, so any observed change of the observed distribution indicates a viewing point bias as the region shifts away from the central meridian.
The main change found in the distribution was a shift of k1 to higher field values.
One reason for this is the larger errors in the line-of-sight magnetic field measurement closer to the solar limb \citep{Hoeksema14}.
These errors are then exaggerated by the cosine correction under the radial field assumption.
Even if there were no errors in the measured field values, the validity of the assumption that the measurements are of a purely radial field decreases for both low field strength 
regions (less vertical magnetic field) and away from the disk centre, where the measured field contains a larger component of the transverse field.
This also introduces errors in the field distribution.

    \idea{Implication for the slope}
A shift in k1 has implications for the slope, with a larger k1 in general giving rise to a steeper slope.
The shift in k1 becomes more pronounced $\sim$ 3 - 4 days from central meridian ($\sim$ 45 - 60 degrees).
In the AR sample analysed in \sects{Obs_Temporal}{Obs_Flux}, if k1 was seen to increase significantly as the region moves away from the central meridian,
then this and any following time steps were removed from the analysis.

    \idea{Bias on ARs away from the central meridian}
To further ensure that the trends seen in \Figs{gradvstime}{gradvsf(flux)} did not result from the position of the AR on the disk in terms of area foreshortening effects and interpolations associated with derotation,
\Fig{gradnearcm} includes only data from ARs within 10 hours, $\sim$ 6 degrees, of their central meridian passage.
The regions were in various stages of their evolution as they passed central meridian. \Fig{gradnearcm} shows the same trends as were obtained with the larger number of data points in \Fig{gradvsf(flux)}, with a slightly more marked difference between the leading and following polarities.
Thus, we conclude that the trends in slope are related to the real field distributions of the ARs.

\begin{figure}
	\centering
	\includegraphics[scale=0.35]{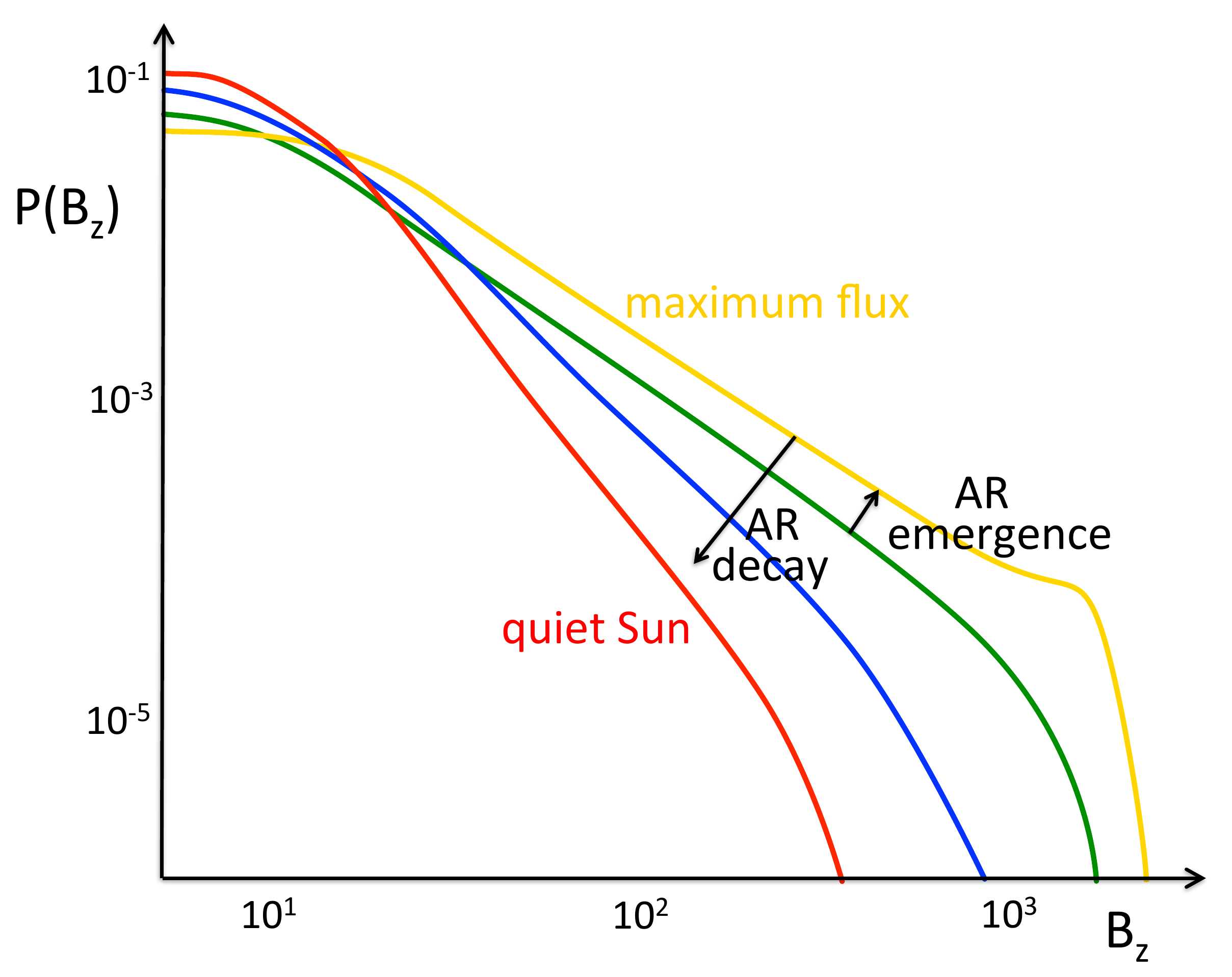}
	\caption{Schema summarising the evolution of the $\Bz$ distribution with a log-log plot. 
		It shows the difference in the distribution between a newly emerging AR (green),  one at around the time of peak flux (yellow) and one during the decay phase (blue).  After this time, the field distribution of the AR evolves towards that of the quiet Sun (red).  
Since both magnetic polarities have very similar distributions at a given time (\eg\ \Figs{example_kde}{decaying_region}) they are not differentiated in this schema.	 }
	\label{fig_prob_evolution}
\end{figure}

\section{Discussion and Conclusions}
\label{sect_Conclusions}

    \idea{Summary of the results}
In this study of 37 emerging ARs, we have shown that there is a relationship between the slope of the vertical component of the photospheric field distribution and the age of an AR. This is summarised in \Fig{prob_evolution}.
At the beginning of a region's emergence, the slope is steep and negative.
The slope becomes less steep, which indicates the coalescence of the fragmented flux that emerges.
Later, the slope reaches a maximum just before the region achieves its peak flux value (at $\sim 0.75-0.8$ peak flux), before the decay processes become dominant.
The slope becomes more negative as the region disperses and this decreasing trend continues towards the quiet Sun slope value of $\sim -3$ (\Fig{decaying_region}).

    \idea{Model the strong field}
A comparison between the observational and theoretical results shows that a simple model of magnetic concentrations can describe the field (flux density) distribution in emerging ARs during the coalescence phase when smaller flux concentrations merge to form larger ones, leading to sunspot formation. 
The model predicts a slope of $\approx -1.67$ for $n=3$, in good agreement with the slope values found in observations of the coalescence phase (\Figs{gradvstime}{gradvsf(flux)}).  

    \idea{Difficulties with the decay}
However, later on there is a major deviation from the classical-diffusion model in the decay phase, indicating that AR magnetic fields do not disperse by simple diffusion.
The latter predicts that after reaching peak flux, the field strength distribution should be characterised by a slope which is evolving from the range $[-1.6,-1.4]$ towards the diffusion exponent value of $-1$.
However, as \Figs{gradvsf(flux)}{gradnearcm} clearly demonstrate, once ARs pass their peak flux and start decaying, their field strength distribution slopes evolve quite differently from these expectations: they start to attain higher negative values. 
Furthermore, ARs measured in the later decay phase display slopes in the range of $[-2.3,-1.6]$, as shown in \Fig{slope_vs_days}, while the quiet Sun, which can be regarded as the end-product of AR decay, shows a slope $\approx -3$. How can we understand this behaviour, which is so clearly opposite to the classical diffusion scenario? 

    \idea{Effect of the convection: proposed scenario}
We suggest that magnetic flux reprocessing by convective cells is responsible for the observed evolution of field distributions. 
Magnetic flux is being gnawed away by granular and supergranular convective cells, in agreement with the turbulent diffusion model \citep[\eg][]{Petrovay97}, which carry away flux concentrations from the strong-field area of ARs. 
\citet{Petrovay97_obs} analysed various theoretical models of sunspot decay using observational data of umbral areas and found that the turbulent diffusion model was the only model supported by the data. The turbulent diffusion model is also consistent with the removal of active region magnetic field by moving magnetic features \citep[\eg][]{Kubo08}, the movement of which is a result of convective flows.
The advected field fragments become concentrated along the boundaries of supergranular cells, where they occasionally meet and cancel with opposite polarity field. The cancelled flux submerges and is re-processed by convection. Part of the reprocessed flux emerges in the centre of supergranular cells as weak intranetwork flux. This process breaks up strong-field flux tubes and makes them emerge as weaker field, effectively transferring strong field to weak field in the field distribution of a decaying AR, making the slope of the field distribution steeper (more negative). The weak intranetwork field is carried to the boundary of the supergranular cells and becomes more concentrated there, which is a counter-mechanism to the former scenario. We propose that the $-3$ slope found in the quiet-Sun field is the result of the combination of these two mechanisms. 

    \idea{Compatibility for global dispersion of ARs?}
Although the evolution of the magnetic field in a decaying AR is very different from what is expected in classical diffusion (slope of $-1$), 
the AR area has previously been found to increase approximately linearly 
with time \citep[\eg][]{vanDriel-Gesztelyi03}, in accordance with the classical diffusion model. Is there a contradiction here? We propose that the increase of AR area is due to the non-stationarity of the supergranular convective cells, which have a lifetime of about one day. This non-stationarity creates a random walk of the flux tubes, which is the underlying physics of classical diffusion.  Therefore the AR evolution can be seen as classical diffusion regarding area increase, while the magnetic field distribution is governed by magneto-convection.  These ideas have to be tested in simulations of emerging ARs that include (or not) magneto-convection.

Other studies \citep[\eg][]{Abramenko10} have characterised the active region magnetic field by its degree of intermittency. It could be interesting to compare this with changes in the field distribution slopes as the resolution of the magnetogram is decreased and particularly for resolutions which do not resolve the highly intermittent small scale structure.

\begin{acknowledgements}
	The authors are grateful to Mark Linton for the interesting discussions and the referee for their insightful comments. 
	We are thankful to the SDO/ HMI and AIA consortia for the data.
	This research has made use of SunPy, an open-source and free community-developed solar data analysis package written in Python \citep{SunPy1}.
	SD and SLY acknowledge STFC for support via their PhD studentships.
	Both DB and LvDG are funded under STFC consolidated grant numbers ST/H00260X/1 and ST/N000722/1.
	LvDG also acknowledges the Hungarian Research grant OTKA K-109276.
	DML is an Early-Career Fellow funded by the Leverhulme Trust.
	DPS is funded by the US Air Force Office of Scientific Research under Grant No. FAQ550-14-1-0213.
	
\end{acknowledgements}

\appendix
\section{Table of ARs}
\label{sect_appendix_table}

37 ARs that emerged on the solar disk between June 2010 and the end of 2014 were studied and are listed in the following table. These ARs emerged in relatively magnetic-field free regions prior to (or around) their central meridian passage.

\begin{table*}[!t]
	\caption{Characteristics of the studied ARs. The first column contains the AR number and the second the time the AR passes the central meridian (CMP).
	The third and fourth columns contain the time difference with CMP of the first appearance and of the maximum unsigned flux, respectively. The fifth column contains the value of the maximum unsigned flux (the average of magnetic flux of both polarities).
	}
	\label{tab_Max}
	\centering
\begin{tabular}{ccccc} 
	\hline \hline
 NOAA &       {Time CMP} &$\Delta t_{\rm start}$&$\Delta t_{\rm max}$& Maximum Flux\\
	  &          [UT]     & [hours] & [hours] & [$10^{21}$ Mx] \\
	\hline
11199 &  2011-04-25 12:00 &   0 &     78 &         6.70 \\
11211 &  2011-05-08 11:00 &  -5 &     31 &         0.73 \\
11242 &  2011-06-28 20:00 & -20 &     46 &         3.89 \\
11273 &  2011-08-17 21:00 & -33 &      9 &         1.42 \\
11284 &  2011-08-26 19:00 & -37 &     47 &         0.65 \\
11398 &  2012-01-13 10:00 & -10 &     68 &         4.64 \\
11437 &  2012-03-19 06:00 & -66 &    -36 &         0.60 \\
11446 &  2012-03-24 00:00 & -30 &     18 &         1.06 \\
11480 &  2012-05-11 07:00 & -43 &     -1 &         0.79 \\
11551 &  2012-08-21 00:00 & -30 &     12 &         1.76 \\
11631 &  2012-12-12 14:00 & -20 &     64 &         5.11 \\
11645 &  2013-01-03 17:00 & -23 &     37 &         2.45 \\
11697 &  2013-03-13 09:00 &   3 &     45 &         1.95 \\
11702 &  2013-03-20 10:00 & -22 &     80 &         3.22 \\
11706 &  2013-03-28 08:00 & -38 &     -2 &         0.82 \\
11707 &  2013-03-31 06:00 & -66 &     24 &         2.26 \\
11709 &  2013-03-28 10:00 & -40 &     20 &         0.94 \\
11750 &  2013-05-15 00:00 &   0 &     84 &         5.06 \\
11765 &  2013-06-07 09:00 & -63 &     75 &         6.40 \\
11768 &  2013-06-11 15:00 &  -3 &     75 &         8.03 \\
11776 &  2013-06-19 08:00 & -20 &     70 &         6.32 \\
11781 &  2013-06-28 17:00 & -23 &     79 &         6.23 \\
11807 &  2013-07-28 17:00 & -11 &     73 &         2.09 \\
11813 &  2013-08-07 12:00 & -30 &     30 &         2.73 \\
11821 &  2013-08-13 07:00 &  -1 &     35 &         1.09 \\
11824 &  2013-08-17 04:00 &  -4 &     50 &         3.05 \\
11825 &  2013-08-18 02:00 &  -8 &     76 &         4.61 \\
11837 &  2013-09-01 17:00 & -29 &     61 &         5.90 \\
11886 &  2013-10-29 10:00 & -34 &     26 &         2.34 \\
11888 &  2013-10-29 17:00 & -53 &     73 &         3.43 \\
11919 &  2013-12-08 09:00 &  15 &     69 &         2.35 \\
11922 &  2013-12-09 21:00 & -33 &     69 &         5.26 \\
11945 &  2014-01-04 09:00 & -57 &     33 &         1.59 \\
12039 &  2014-04-16 20:00 & -26 &     40 &         1.17 \\
12089 &  2014-06-13 06:00 & -60 &     84 &        12.40~~ \\
12098 &  2014-06-26 10:00 & -64 &      2 &         2.41 \\
12119 &  2014-07-20 04:00 & -46 &     26 &         2.59 \\
\hline
\end{tabular}
\end{table*}


\bibliographystyle{aa}
\bibliography{My_Collection}

\end{document}